\documentclass[12pt]{article}
\usepackage[latin1]{inputenc}

\usepackage{amsmath}
\usepackage{color}
\usepackage{amsfonts}
\usepackage{amssymb}
\usepackage{graphicx}
\usepackage{geometry}
\usepackage{amssymb,epsfig,subfigure}
\usepackage{hyperref}
\usepackage{comment}
\usepackage{tabularx}
\usepackage{bm}
\usepackage{euscript}
\usepackage{graphicx}
\usepackage{color}
\usepackage{amsfonts}
\usepackage{exscale}
\usepackage{amsbsy}
\usepackage{subfigure}
\usepackage{textcomp}
\usepackage{comment}
\usepackage{hyperref}
\usepackage{slashed}
\usepackage{authblk}

\usepackage{tabularx}
\usepackage{bm}
\usepackage{euscript}
\usepackage{graphicx}
\usepackage{color}
\usepackage{amsfonts}
\usepackage{exscale}
\usepackage{amsbsy}
\usepackage{subfigure}
\usepackage{textcomp}
\usepackage{comment}
\usepackage{hyperref}

\usepackage[numbers,sort&compress]{natbib}

\def\simge{%  ``greater than about'' symbol
    \mathrel{\rlap{\raise 0.511ex 
        \hbox{$>$}}{\lower 0.511ex \hbox{$\sim$}}}}

\def\simle{%  ``less than about'' symbol
    \mathrel{\rlap{\raise 0.511ex 
        \hbox{$<$}}{\lower 0.511ex \hbox{$\sim$}}}}

\makeatletter
\renewcommand\section{\@startsection {section}{1}{\z@}%
                                 {-3.5ex \@plus -1ex \@minus -.2ex}%nn
                                   {2.3ex \@plus.2ex}%
                                   {\normalfont\large\bfseries}}
\renewcommand\subsection{\@startsection{subsection}{2}{\z@}%
                                   {-3.25ex\@plus -1ex \@minus -.2ex}%
                                     {1.5ex \@plus .2ex}%
                                     {\normalfont\bfseries}}
\renewcommand\subsubsection{\@startsection{subsubsection}{3}{\z@}%
                                   {-3.25ex\@plus -1ex \@minus -.2ex}%
                                     {1.5ex \@plus .2ex}%
                                     {\normalfont\itshape}}
\makeatother

\def\pplogo{\vbox{\kern-\headheight\kern -29pt
\halign{##&##\hfil\cr&{\ppnumber}\cr\rule{0pt}{2.5ex}&\ppdate\cr}}}
\makeatletter
\def\ps@firstpage{\ps@empty \def\@oddhead{\hss\pplogo}%
  \let\@evenhead\@oddhead % in case an article starts on a left-hand page
}%      The only change in \maketitle is \thispagestyle{firstpage} instead of 
\def\maketitle{\par
 \begingroup
 \def\thefootnote{\fnsymbol{footnote}}
 \def\@makefnmark{\hbox{$^{\@thefnmark}$\hss}}
 \if@twocolumn
 \twocolumn[\@maketitle]
 \else \newpage
 \global\@topnum\z@ \@maketitle \fi\thispagestyle{firstpage}\@thanks
 \endgroup
 \setcounter{footnote}{0}
 \let\maketitle\relax
 \let\@maketitle\relax
 \gdef\@thanks{}\gdef\@author{}\gdef\@title{}\let\thanks\relax}
\makeatother

\numberwithin{equation}{section}

\renewcommand{\dag}{\dagger}

\newcommand{\be}{\begin{equation}}
\newcommand{\bea}{\begin{eqnarray}}
\newcommand{\ee}{\end{equation}}
\newcommand{\eea}{\end{eqnarray}}
\newcommand\beq{\begin{equation}}
\newcommand\eeq{\end{equation}}

\newcommand{\mc}{\mathcal}

\renewcommand{\t}{\tilde}

\newcommand{\sgn}{{\rm{sgn}}}

\def\be{\begin{equation}}
\def\ee{\end{equation}}
\def\ba#1\ea{\begin{align}#1\end{align}}
\def\bg#1\eg{\begin{gather}#1\end{gather}}
\def\bm#1\em{\begin{multline}#1\end{multline}}
\def\bmd#1\emd{\begin{multlined}#1\end{multlined}}

\def\({\left(}
\def\){\right)}
\def\[{\left[}
\def\]{\right]}

\begin{document}

\setcounter{page}0
\def\ppnumber{\vbox{\baselineskip14pt
%\hbox{hep-th/0000000}
}}

\def\ppdate{%\footnotesize{SU/ITP-14/XX}
} \date{}

%\author{Shamit Kachru$^{1}$, Michael Mulligan$^{2,1}$, Gonzalo Torroba$^3$, Huajia Wang$^4$\\
%[7mm] \\
%{\normalsize \it $^{1}$Stanford Institute for Theoretical Physics, Stanford 
%University, Stanford, CA 94305, USA} \\
%{\normalsize \it $^3$Department of Physics and Astronomy, University of California, Riverside, California 92511, USA}\\
%{\normalsize \it $^3$Centro At\'omico Bariloche and CONICET, R8402AGP Bariloche, 
%Argentina}\\
%{\normalsize \it $^4$ Department of Physics, University of Illinois, Urbana-Champaign, Illinois 61801, USA}
%}

\title{\bf  Bosonization and Mirror Symmetry
\vskip 0.5cm}
\author[1]{Shamit Kachru}
\author[2,1]{Michael Mulligan}
\author[3]{Gonzalo Torroba}
\author[4]{Huajia Wang}
\affil[1]{\small \it Stanford Institute for Theoretical Physics, Stanford University, Stanford, CA 94305, USA}
\affil[2]{\small \it Department of Physics and Astronomy, University of California,
Riverside, CA 92511, USA}
\affil[3]{\small \it Centro At\'omico Bariloche and CONICET, R8402AGP Bariloche, ARG}
\affil[4]{\small \it Department of Physics, University of Illinois, Urbana-Champaign, IL 61801, USA}

\bigskip

\maketitle

\begin{abstract}
We study bosonization in 2+1 dimensions using mirror symmetry, a duality that relates pairs of supersymmetric theories. 
Upon breaking supersymmetry in a controlled way, we dynamically obtain the bosonization duality that equates the theory of a free Dirac fermion to QED3 with a single scalar boson. This duality may be used to demonstrate the bosonization duality relating an $O(2)$-symmetric Wilson-Fisher fixed point to QED3 with a single 
Dirac fermion, Peskin-Dasgupta-Halperin duality, and the recently conjectured duality relating the theory of a free Dirac fermion to fermionic QED3 with a single flavor. Chern-Simons and BF couplings for both dynamical and background gauge fields play a central role in our approach. In the course of our study, we describe a ``chiral'' mirror pair that may be viewed as the minimal supersymmetric generalization of the two bosonization dualities.
\end{abstract}
\bigskip

\newpage

\tableofcontents

\vskip 1cm

%%%%%%%%%%%%%%%%%%%%%%%%%%%%%%%%%%%%%%%%%%%%%%%%%%%%%%%%%%%%%%%%%
%%%%%%%%%%%%%%%%%%%%%%%%%%%%%%%%%%%%%%%%%%%%%%%%%%%%%%%%%%%%%%%%%
%%%%%%%%%%%%%%%%%%%%%%%%%%%%%%%%%%%%%%%%%%%%%%%%%%%%%%%%%%%%%%%%%
%%%%%%%%%%%%%%%%%%%%%%%%%%%%%%%%%%%%%%%%%%%%%%%%%%%%%%%%%%%%%%%%%
\section{Introduction}\label{sec:intro}

Bosonization is a duality that equates a fermionic description of a particular system to an alternative bosonic one.
The classic example -- which occurs in two spacetime dimensions (1+1D) --  relates a self-interacting Dirac fermion to a scalar boson with cosine potential \cite{Colemansinegordon1974, LutherPeschel1974, LutherEmery1974, Mandelstam1975}.
The direct demonstration for the duality constructs the Dirac fermion from a coherent state of bosons \cite{Mandelstam1975}.
%A transformation that maps the Dirac fermion into a coherent state of bosons provides a direct demonstration for the equivalence. 
This duality has had incredible utility for the description of 1+1D condensed matter systems that range from spin models and itinerant fermions to the excitations living on the edges of quantum Hall droplets \cite{Giamarchibook, Fradkinbook}.

The situation is different in 2+1D where various (physically-motivated) bosonization proposals are not yet rigorously established, in the sense of \cite{Mandelstam1975}, despite their successful application to a variety of condensed matter systems \cite{Fradkinbook}.
Recently, there has been substantial progress in motivating a large class of new bosonization dualities \cite{GMPTWY2012, AharonyGurAriYacoby2012, AharonyGurAriYacobysecond2012}. 
Aharony \cite{Aharony2016} (see also \cite{Radicevic2016, HsinSeiberg2016}) has clarified the basic structure of these conjectured dualities (indicated by $\leftrightarrow$):
\begin{align}
\label{dualityone}
& {\rm N_f\ fermions\ coupled\ to\ } SU(k)_{-N + {N_f \over 2}} \leftrightarrow {\rm N_f\ scalars\ coupled\ to\ } U(N)_{k,k}; \\
\label{dualitytwo}
& {\rm N_f\ scalars\ coupled\ to\ } SU(N)_{k} \leftrightarrow {\rm N_f\ fermions\ coupled\ to\ } U(k)_{-N + {N_f \over 2}, -N + {N_f \over 2}}; \\
\label{dualitythree}
& {\rm N_f\ fermions\ coupled\ to\ } U(k)_{-N + {N_f \over 2}, - N \mp k + {N_f \over 2}} \leftrightarrow {\rm N_f\ scalars\ coupled\ to\ } U(N)_{k, k \pm N}.
\end{align}
The two-component Dirac fermions and scalar bosons transform in the fundamental representation of the gauge group. 
The subscripts give the levels of Chern-Simons terms with $U(N)_{k, l} \equiv (SU(N)_k \times U(1)_{Nl})/{\mathbb{Z}_N}$.
\eqref{dualityone} - \eqref{dualitythree} have been validated in the large $N$ 't Hooft limit in which the ratio $N/k$ is held fixed \cite{GMPTWY2012, AharonyGurAriYacoby2012, AharonyGurAriYacobysecond2012}.
At finite $N$, evidence has come in the form of consistency checks wherein conjectured dual pairs have matching phase structure \cite{AharonyGurAriYacobysecond2012} or may be obtained upon deformation of better-understood supersymmetric (SUSY) parent theories \cite{JainMinwallaYokoyama2013, Gur-AriYacoby2015}.

In this paper, we derive the $N_f = N = k = 1$ versions of \eqref{dualityone} and \eqref{dualitytwo} and find that they are realized via the 2+1D effective lagrangians,\footnote{Explanation of the precise meaning of the level-1/2 Chern-Simons terms is provided in \S \ref{subsec:superspace}.}
\begin{align}
\label{firstduality}
& \bar{\Psi} i \slashed{D}_{\hat{A}} \Psi - {1 \over 8 \pi} \hat{A} d \hat{A} \leftrightarrow |D_{-a} \varphi|^2 - |\varphi|^4 + {1 \over 4 \pi} a d a - {1 \over 2 \pi} \hat{A} d a, \\
\label{secondduality}
& |D_{\hat{A}} \phi|^2 - |\phi|^4 + {1 \over 4 \pi} \hat{A} d \hat{A} \leftrightarrow \bar{\psi} i \slashed{D}_{a} \psi - {1 \over 8 \pi} a d a - {1 \over 2 \pi} \hat{A} d a.
\end{align}
In the above relations, $\hat{A}$ represents a background $U(1)$ gauge field, while $a$ is a dynamical 2+1D $U(1)$ gauge field.\footnote{Our conventions for writing Chern-Simons and BF terms for gauge fields $A = A_\mu$ and $B = B_\mu$ is the following: $A d B \equiv \epsilon^{\mu \nu \rho} A_\mu \partial_\nu B_\rho$ with $\mu,\nu,\rho \in \{t,x,y\}$ and $\epsilon^{txy} = 1$.
The covariant derivative with respect to $\pm A$ is denoted by $D_{\pm A} \equiv \partial_\mu \mp i A_\mu$.
Hats are used to indicate background fields.}
\eqref{firstduality} relates a two-component Dirac fermion to three-dimensional quantum electrodynamics (QED3) with a single scalar boson and a level-1 Chern-Simons term for the dynamical gauge field.
The left-hand side of \eqref{secondduality} is simply the $O(2)$-symmetric Wilson-Fisher critical point, while the right-hand side is QED3 with a single Dirac fermion and a level-$1/2$ Chern-Simons term for the dynamical gauge field.
In both dualities, there are important Chern-Simons terms for and BF couplings to $\hat{A}$ that ensure their validity.
Prior work studying proposals closely related to \eqref{firstduality} and \eqref{secondduality} includes \cite{ChenFisherWu1993, wenwu1993, Fradkin:1996xb, BarkeshliMcGreevy2012continuous}.

Our approach to establishing \eqref{firstduality} and \eqref{secondduality} is to deform the SUSY duality known as mirror symmetry \cite{Intriligator:1996ex, deBoer:1996mp, deBoeroz, deBoeroztwo, Aharony, Kapustin:1999ha}. This is motivated by our previous work \cite{Kachru:2015rma} which used mirror symmetry to relate the half-filled Landau level with two flavors to a composite fermion theory with an emergent gauge field.
We focus on the simplest example that equates a free ${\cal N} = 4$ hypermultiplet -- {\it theory A} -- to a ${\cal N}=4$ hypermultiplet interacting via a ${\cal N}=4$ vector multiplet with $U(1)$ gauge group -- {\it theory B}.
Our first step in \S \ref{sec:overview} is to review this duality and show how to map various ${\cal N}=2$ SUSY-preserving deformations across the duality.
These deformations enable us to show in \S \ref{sec:chiralmirror} the equivalence of a single free ${\cal N}=2$ chiral multiplet and ${\cal N}=2$ SUSY QED3 with a single chiral multiplet as first obtained in \cite{Tong:2000ky}.
The chiral duality of \S \ref{sec:chiralmirror} provides a minimal SUSY generalization of \eqref{firstduality} and \eqref{secondduality}.

In \S \ref{bosonizationarguments}, we present the main result of the work: we show that a specific SUSY-breaking perturbation of the chiral duality results in \eqref{firstduality}.
Because theory A is free, the effects of the deformations we consider are easily understood: we show that there exist two distinct massive phases in a particular parameter regime that are separated by a single critical point whose lagrangian description is that of a free Dirac fermion, i.e., the left-hand side of \eqref{firstduality}.
Mirror symmetry dictates identical phase structure as parameters are varied in theory B: there must be a single critical point (within the neighborhood of variations we consider) and a matching of the effective actions for various background gauge fields in the nearby massive phases.\footnote{More precisely, duality requires that the differences of the theory A and theory B effective actions across the phase transition must match. In this way, regularization-dependent counterterms cancel out.}  
These two requirements uniquely constrain what field must become light at the critical point in the theory B description and allow us to deduce the right-hand side of \eqref{firstduality}.
Our arguments are rather general and help us temper the interesting, but subtle, strong dynamics of the theory B description.

We note that this approach is purely 2+1D in nature.
We do not consider theories arising on surfaces of higher-dimensional spaces, e.g., surface phases of a topological insulator, whose proper regularization is provided by the higher-dimensional bulk \cite{MulliganBurnell2013, Wittenfermionpathintegrals2016}.

The derivation of this bosonization duality has far-reaching consequences.
As recent work \cite{Karch:2016sxi, Seiberg:2016gmd, 2016arXiv160601912M, HsinSeiberg2016} has shown, if \eqref{firstduality} (or \eqref{secondduality}) is assumed, a large web of dualities can be found through a sequence of modular transformations \cite{WittenSL2Z2003, LeighPetkouSL2Z2003}. In particular, the
Peskin-Dasgupta-Halperin duality \cite{Peskin:1977kp, DasguptaHalperin1981},
\be\label{eq:PDH}
|D_{\hat{A}} \phi|^2 - |\phi|^4 \; \leftrightarrow \; |D_a \varphi|^2 - |\varphi|^4 - {1 \over 2 \pi} \hat{A} d a,
\ee
and the ``topological completion'' of a recent ``fermion/fermion" duality conjecture \cite{Son2015, WangSenthilfirst2015, MetlitskiVishwanath2016}
\be\label{eq:son}
\bar{\Psi} i \slashed{D}_{\hat{A}} \Psi - {1 \over 8 \pi} \hat{A} d \hat{A} \; \leftrightarrow\; \bar{\Psi} i \slashed{D}_{-a} \Psi + {1 \over 8 \pi} a d a + {1 \over 2\pi} b d a + {2 \over 4 \pi} b d b - {1 \over 2 \pi} \hat{A} d b\,.
\ee
follow from \eqref{firstduality}. Here, as before, $\hat{A}$ is a background $U(1)$ gauge field, while $a$ and $b$ are dynamical 2+1D $U(1)$ gauge fields.

Son's conjecture \cite{Son2015} and its extension to a general fermion/fermion duality conjecture by Metlitski, Senthil, Vishwanath, and Wang \cite{WangSenthilfirst2015, MetlitskiVishwanath2016} has become the subject of and inspiration for much recent activity in condensed matter physics \cite{Kachru:2015rma, Geraedtsetal2015, Metlitski:2015yqa, XuYou2015selfdual, PotterSerbynVishwanath2015, MurthyShankar2016halfull, MrossAliceaMotrunichexplicitderivation2016, WangSenthilsecond2016, MulliganRaghuFisher2016, WangSenthil2016, BalramJain2016, MrossAliceaMotrunichbosonicph2016, Mulligan2016, WangChakravarty2016, MilovanovicDimitrijevicjuricic2016}.
In short, these works introduce a manifestly ``particle-hole" symmetric\footnote{The particle-hole transformation allows a perturbative description of the lowest-Landau level using either the empty electron vacuum or the empty hole vacuum, i.e., the filled Landau level. 
Particle-hole symmetry is an emergent symmetry of the actual physical system that may occur at half-filling of the lowest-Landau level (when the electron density is precisely half the value of the applied magnetic field) where these two descriptions may become equivalent. 
Son's conjecture provides a manifestly particle-hole symmetric starting point for any such description -- something not easily achieved \cite{kivelson1997, DHLee1998, BMF2015} using the conventional approach pioneered by Halperin, Lee, and Read \cite{halperin1993, Kalmeyer1992}.
The extension \cite{WangSenthilfirst2015, MetlitskiVishwanath2016} states that this fermion/fermion duality conjecture continues to hold in vanishing magnetic field.} description for the half-filled lowest Landau level of the 2+1D electron gas \cite{Son2015} and a dual description \cite{WangSenthilfirst2015, MetlitskiVishwanath2016} for the time-reversal invariant Dirac surface state of a 3+1D topological insulator \cite{FuKaneMele, MooreBalents, Roy2009, qhz2008}.
We hope that our derivation based on mirror symmetry provides additional insight into these problems and related ones where such dualities are useful.

The remainder of this paper is organized as follows.
In \S \ref{sec:overview}, we review ${\cal N}=4$ mirror symmetry and show how a certain class of ${\cal N}=2$ preserving deformations map across the duality.
In \S \ref{sec:chiralmirror}, we demonstrate the SUSY duality between a free ${\cal N}=2$ chiral multiplet and a ${\cal N}=2$ vector multiplet coupled to a ${\cal N}=2$ chiral multiplet.
Finally, we derive \eqref{firstduality} in \S \ref{bosonizationarguments}.
We conclude in \S \ref{sec:concl} and outline possible directions of future work.
The reader interested in avoiding SUSY notation (which is explained) can jump straightaway to \S \ref{bosonizationarguments}.
Appendix \ref{app:superspace} reviews the basics of superspace in 2+1D.

%%%%%%%%%%%%%%%%%%%%%%%%%%%%%%%%%%%%%%%%%%%%%%%%%%%%%
%%%%%%%%%%%%%%%%%%%%%%%%%%%%%%%%%%%%%%%%%%%%%%%%%%%%%
%%%%%%%%%%%%%%%%%%%%%%%%%%%%%%%%%%%%%%%%%%%%%%%%%%%%%
%%%%%%%%%%%%%%%%%%%%%%%%%%%%%%%%%%%%%%%%%%%%%%%%%%%%%
\section{Mirror symmetry and its deformations}\label{sec:overview}

The aim of this section is twofold. 
First, we will review the basic tools of SUSY that we will use to derive the bosonization duality \eqref{firstduality}. 
These tools include the SUSY dualities between certain 2+1D theories known as mirror symmetry \cite{Intriligator:1996ex, deBoer:1996mp, deBoeroz, deBoeroztwo, Aharony, Kapustin:1999ha}.\footnote{More recent developments may be found in \cite{BorokhovKapustinWumirrormonopoles, KapustinWillettYaakov}.}
We focus on one example of this duality (which we refer to as mirror symmetry for convenience).

Our second goal is to extend mirror symmetry to include deformations by background superfields that couple to the ``non-topological" global currents. 
The role of the ``topological" $U(1)_J$ symmetry (reviewed below) was already understood in the first works on the subject -- see e.g. \cite{Kapustin:1999ha}. 
However, the mirror theories contain additional global symmetries (an axial symmetry and an R-symmetry); we will explain how the deformations associated to these symmetries are mapped across the duality.
This mapping will be the central ingredient in our approach. 
In the remainder of the paper, we will show that within certain parameter regimes of the backgrounds fields that maintain Lorentz invariance, but may break SUSY, the ${\cal N}=4$ mirror duality can be shown to either flow to a ``chiral" ${\cal N}=2$ duality or the non-SUSY bosonization duality \eqref{firstduality}. 
%that has been conjectured \cite{Aharony2016} and recently argued to hold~\cite{Karch:2016sxi, Seiberg:2016gmd, HsinSeiberg2016}.

\subsection{Superfields and lagrangians}\label{subsec:superspace}

Let us begin with a quick review of superfields, their components, and interactions. More details about superspace are given in Appendix \ref{app:superspace}.

A simple formulation of superspace in 2+1D obtains by starting from $\mc N=1$ superspace (i.e., four supercharges) in 3+1D, and dimensionally reducing along the $x^2$ direction. 
The resulting $\mc N=2$ superspace has the following two basic superfields. 
A chiral superfield $\Phi$ that is composed of a complex scalar $\phi$, a two-component Dirac fermion $\psi$, and an auxiliary complex field $F$. A vector superfield $V$ that contains a gauge field $A_\mu$, a real scalar $\sigma$ (which can be thought as the component of the 3+1D gauge field along the reduced dimension), a gaugino (two-component Dirac fermion) $\lambda$, and an auxiliary real field $D$.

Their lagrangians can be compactly written in superspace. The kinetic term for a chiral superfield $\Phi$ of charge $q$ under the $U(1)$ symmetry gauged by the vector superfield $V$ is
\bea
\mc L^C(\Phi, V)& =& \int d^4 \theta\,\Phi^\dag e^{2q V} \Phi \\
&=&|D_{q A} \phi|^2+ \bar \psi i \slashed{D}_{q A} \psi-(q \sigma)^2 |\phi|^2-q \sigma \bar \psi \psi-i q ( \phi^*  \lambda \psi - \phi \bar \psi \bar \lambda)- q D |\phi|^2\nonumber\,.
\eea
The covariant derivative $D_{q A} \equiv \partial_\mu -i q A_\mu$ and $\slashed{D}_{q A} \equiv \gamma^\mu (\partial_\mu -i q A_\mu)$ with $\mu = 0,1,2$. 
The kinetic term for a vector superfield is
\be
\mc L^V(V) = \frac{1}{4g^2} \int d^2 \theta\,W_\alpha^2 +\text{h.c.}=\frac{1}{g^2}\left(-\frac{1}{4} F_{\mu\nu}^2+\frac{1}{2}(\partial \sigma)^2 + \bar \lambda i \slashed{\partial} \lambda+\frac{1}{2}D^2\right)\,,
\ee
where the field strength $F_{\mu \nu} = \partial_\mu A_\nu - \partial_\nu A_\mu$.
We use the Lorentzian-signature metric $\eta^{\mu\nu}=\text{diag}(+,-,-)$ and work with gamma matrices satisfying
\be
\label{gammaconventions}
\lbrace \gamma^\mu, \gamma^\nu \rbrace= 2 \eta^{\mu\nu}\;\;,\;\;(\gamma^0 \gamma^1 \gamma^2)_{\alpha \beta}=-i \delta_{\alpha \beta}\,.
\ee
As an example, we may take the representation:
\begin{align}
\gamma^0 = \sigma^3\,,\; \gamma^1 = i \sigma^1\,,\; \gamma^2 = i \sigma^2\,.
\end{align}
(We note that this is not the representation that obtains from the dimensional reduction given in Appendix \ref{app:superspace}.)

Chern-Simons and BF terms will also appear in our dualities, so let us review their $\mc N=2$ version~\cite{Ivanov:1991fn, Gaiotto:2007qi}. The $\mc N=2$ BF coupling between two vector multiplets $V^{(1)}$ and $V^{(2)}$ is
\bea\label{eq:LBF-general}
\mc L_{BF}^{\mc N=2}(V^{(1)}, V^{(2)})&=& \frac{1}{2\pi} \int d^4 \theta\, V^{(1)}\, \Sigma^{(2)}\\
&=& \frac{1}{2\pi}\left( \epsilon^{\mu\nu \rho} A^{(1)}_\mu \partial_\nu A^{(2)}_\rho+ D^{(1)} \sigma^{(2)}+D^{(2)} \sigma^{(1)}+ \frac{1}{2}(\bar \lambda^{(1)}\lambda^{(2)}+\bar \lambda^{(2)}\lambda^{(1)})\right)\nonumber\,,
\eea
where $\Sigma= \bar D^\alpha D_\alpha V$ (the superspace derivative $D_\alpha$ is defined in Appendix \ref{app:superspace}). In this notation, a SUSY Chern-Simons lagrangian at level $k$ corresponds to
\be
\label{N2CS}
\mc L_{CS}^{\mc N=2}(V)=\mc L_{BF}^{\mc N=2}(V, V) = \frac{k}{4\pi} \int d^4 \theta\, V \Sigma=\frac{k}{4\pi} \left(\epsilon^{\mu\nu\rho} A_\mu \partial_\nu A_\rho+2 D \sigma + \bar \lambda \lambda \right)\,.
\ee
A one-loop calculation shows that integrating out chiral superfields $\Phi_f$ of mass $m_f$ and charge $q_i^f$ under $U(1)_i$ produces a $\mc N=2$ Chern-Simons term:
\be\label{eq:loopCS}
\mc L_{CS}^{\mc N=2}=\frac{k_{ij}}{4\pi}\int\,d^4 \theta\, V_i \,\Sigma_j\;\;,\;\;k_{ij}=\frac{1}{2}\sum_f\, q_i^f q_j^f\,\sgn(m_f)\,.
\ee
Our convention for the fermion mass sign is $\mc L \supset - m_f \bar \psi_f \psi_f$.

Let us comment on a subtle point regarding Chern-Simons terms generated by integrating out a single fermion. This wil also clarify the statement of the dualities in \eqref{firstduality} and \eqref{secondduality} -- see \cite{AlvarezGaume:1984nf, Wittenfermionpathintegrals2016, Seiberg:2016gmd, HsinSeiberg2016} for further discussion.\footnote{We thank N. Seiberg, T. Senthil, and C. Wang for correspondence on this point.}
When a Dirac fermion of mass $m$ is integrated out, the effective action obtains the correction $\delta S = {\pi {\rm sgn}(m) \over 2} \eta(A,g)$, where $\eta(A,g)$ is the eta-invariant and $A$ and $g$ are the gauge field and metric to which the fermion couples.
We will be exclusively interested in a setting in which the background metric is flat Minkowski space $g = \delta_{2,1}$ so we will not discuss the metric contribution to $\eta(A,g)$.
In our expressions, we substitute ${\pi \over 2} \eta(A, \delta_{2,1}) = {1 \over 8 \pi} \int d^3x\ A d A$ as short-hand; in general, this equality is only true mod $\pi \mathbb{Z}$ (see Eq. (2.50 of \cite{Wittenfermionpathintegrals2016}).
Thus, in writing Chern-Simons terms in this paper, it is to be understood that we have chosen, e.g., a time-reversal invariant Pauli-Villars regularization of our theories, which in the UV contain an even number of Dirac fermions; without such a specification, a correct statement requires the eta-invariant.

We now have all the necessary ingredients to discuss $\mc N=4$ SUSY theories. The two multiplets that will be relevant to us are the $\mc N=4$ hypermultiplet $\mc U$, which contains two $\mc N=2$ chiral multiplets $U_+$ and $U_-$, and the
$\mc N=4$ vector multiplet $\mc V$, which contains an $\mc N=2$ vector multiplet $V$ and chiral multiplet $\Phi$. The hypermultiplet lagrangian for a charged superfield is
\bea
\mc L^{\mc H}(\mc U, \mc V)&=&\int d^4 \theta\,(U_+^\dag e^{2 V} U_++U_-^\dag e^{-2 V} U_-)+ \int d^2 \theta\,i \sqrt{2} \Phi U_+ U_-+\text{h.c.}\\
&=& |D_{\pm a}u_\pm|^2 + \bar i \psi_\pm \not \! \! D_{\pm a} \psi_\pm - (\sigma^2+|\phi|^2)(|u_+|^2+|u_-|^2)-D (|u_+|^2-|u_-|^2)+ F u_+ u_- \nonumber\\
&-& \sigma (\bar \psi_+ \psi_+- \bar \psi_- \psi_-) - \phi \psi_+ \psi_- -i \psi_\phi(u_+ \psi_-+ u_- \psi_+)- i\lambda(u_+^\dag \psi_+- u_-^\dag \psi_-)+\text{h.c.}\nonumber
\eea
The vector-multiplet lagrangian is
\bea
\mc L^{\mc V}(\mc V)&=& \frac{1}{4g^2} \int d^2 \theta\,W_\alpha^2 +\text{h.c.}+ \frac{1}{g^2} \int d^4 \theta\,\Phi^\dag \Phi \\
&=&\frac{1}{g^2}\left(-\frac{1}{4} f_{\mu\nu}^2+\frac{1}{2}(\partial \sigma)^2 + |\partial \phi|^2+ \bar \lambda i \slashed{\partial} \lambda+ \bar \psi_\phi i \slashed{\partial} \psi_\phi+\frac{1}{2}D^2+|F|^2\right) \nonumber\,,
\eea
where $f_{\mu \nu} = \partial_\mu a_\nu - \partial_\nu a_\mu$.
Finally, the $\mc N=4$ version of the BF coupling is
\be
\mc L_{BF}^{\mc N=4}(\mc V^{(1)},\mc V^{(2)})=\frac{1}{2\pi} \int d^4 \theta\, V^{(1)}\, \Sigma^{(2)}- \frac{1}{2\pi} \int d^2 \theta\, \Phi^{(1)} \Phi^{(2)}+\text{h.c.} 
\ee
As the above expressions make clear, superspace allows quite simple and compact forms for component lagrangians that may seem rather involved.
%We should stress that although these component lagrangians may seem rather involved, the expressions in terms of superspace variables are quite simple.

%%%%%%%%%%%%%%%%%%%%%%%%%%%%%%%%%%%%%%%%%%%%%%%%%%%%%%
%%%%%%%%%%%%%%%%%%%%%%%%%%%%%%%%%%%%%%%%%%%%%%%%%%%%%%
\subsection{$\mc N=4$ mirror symmetry}

The simplest example of $\mc N=4$ mirror symmetry is a duality between the following two theories. %(We follow closely the approach of~\cite{Kapustin:1999ha}.)
We focus exclusively on this particular example.

{\it Theory A} is the theory of a free hypermultiplet $\mc Q$.
In $\mc N=2$ notation, this is given by two chiral multiplets $(V_+, V_-)$, each of which contains a complex scalar $v_\pm$ and a two-component Dirac fermion $\Psi_\pm$. A crucial role will be played by a $U(1)_J$ global symmetry, under which the supermultiplets $V_\pm$ have charges $\pm 1$. The theory has nonabelian $SU(2)_L \times SU(2)_N$ ``R-symmetries," under which $(v_+, v_-^*)$ and $(\Psi_+, \Psi_-^*)$ transform as $(2,1)$ and $(1,2)$, respectively. 
The field content and charges are summarized in (\ref{tab:thA}).
\begin{center}
\be\label{tab:thA}
\begin{tabular}{c|ccc}
&$SU(2)_L$&$SU(2)_N$&$U(1)_J$\\
\hline
&&&\\[-12pt]
$v_i \equiv (v_+, v_-^*)$  & $2$ & 1 & 1  \\
&&&\\[-12pt]
$\Psi_a \equiv (\Psi_+, \Psi_-^*)$  & 1& $2$ & 1
\end{tabular}
\ee
\end{center}
Because theory A is free, these symmetries are exact.
(The R-symmetries do not commute with SUSY since $SU(2)_L$ and $SU(2)_N$ act separately on the bosons and fermions.)

The Cartan subgroup of the global symmetry is $U(1)_L \times U(1)_N \times U(1)_J$.
It will be convenient to consider the following linear combination of symmetries: $U(1)_R \equiv U(1)_L$ and $U(1)_A \equiv U(1)_N - U(1)_L$ (the linear combination $U(1)_A$ commutes with SUSY).
The charge assignments for the fields under $U(1)_R \times U(1)_A \times U(1)_J$ are given in \eqref{eq:thAchargescartan}.
\be\label{eq:thAchargescartan}
\begin{tabular}{c|ccc}
& $U(1)_R$ & $U(1)_A$ & $U(1)_J$ \\
\hline
&&&\\[-12pt]
$v_+$ & 1 & -1 & 1 \\
&&&\\[-12pt]
$v_-$ & 1 & -1 & -1 \\
&&&\\[-12pt]
$\Psi_+$ & 0 & -1 & 1 \\
&&&\\[-12pt]
$\Psi_-$ & 0 & -1 & -1
\end{tabular}
\ee

In the presence of a background ${\cal N}=4$ vector superfield\footnote{We denote background non-dynamical fields with `hats'.} $\hat{ \mc  V}_J$ for the $U(1)_J$ symmetry, the lagrangian of theory A is
\be\label{eq:LA}
\mc L^{(A)}(\mc Q, \hat{ \mc V}_J) = \mc L^{\mc H}(\mc Q, \hat{\mc V}_J)=\int d^4 \theta \left( V_+^\dag e^{2 \hat  V_J} V_++V_-^\dag e^{-2 \hat  V_J} V_-\right)+ \int d^2 \theta\,\sqrt{2} i \hat \Phi_J V_+ V_- +\text{h.c.}
\ee
This defines a partition function
\be
Z^{(A)}[\hat {\mc V}_J]= \int D{ \mc Q} \,\exp\left(i\int d^3x\, \mc L^{(A)}(\mc Q, \hat{\mc V}_J)\right)\,.
\ee

{\it Theory B} is ${\cal N}=4$ SUSY QED3 with a single charged hypermultiplet.
Mirror symmetry says that this theory has the non-interacting description provided by theory A. 
Our notation for the matter content of theory B is as follows. The $\mc N=4$ vector multiplet contains a $\mc N=2$ vector multiplet $V=(a_\mu, \sigma, \lambda)$ and a $\mc N=2$ neutral chiral multiplet $\Phi= (\phi, \psi_\phi)$. Here $\sigma$ is a real scalar, $\phi$ is a complex scalar, and $\lambda$ and $\psi_\phi$ are two-component Dirac fermions. The $\mc N=4$ charged hypermultiplet contains $\mc N =2$ chiral multiplets $U_\pm = (u_\pm, \psi_\pm)$ of opposite charge under $a_\mu$.  The $U(1)_J $ global symmetry of theory B arises from dualizing the field strength,
\be\label{eq:Jmu2}
J_\mu=\frac{1}{2\pi} \epsilon_{\mu\nu\rho} \partial^\nu a^\rho\,,
\ee
whose conservation law is equivalent to the Bianchi identity for the emergent gauge field. It acts as a shift on the dual photon $\gamma$, where $f_{\mu\nu}=\partial_\mu a_\nu-\partial_\nu a_\mu=\epsilon_{\mu\nu\rho} \partial^\rho \gamma$. 
Mirror symmetry identifies the global symmetries of both theories.
The gauge field then arises from dualizing the $U(1)_J$ current of theory A.
The rest of the fields are {\it neutral} under $U(1)_J $. On the other hand, the symmetries $SU(2)_L \times SU(2)_N$ act as $(3,1)$ on the triplet of scalars $(\sigma, \phi)$, $\lambda, \psi_\phi$ are in the bifundamental, $(u_+, u_-^*)$ transform as $(1,2)$, and $(\psi_+, \psi_-^*)$ are in the $(2,1)$. This is summarized in (\ref{tab:thB}).
\begin{center}
\be\label{tab:thB}
\begin{tabular}{c|ccc}
&$SU(2)_L$&$SU(2)_N$&$U(1)_J$\\
\hline
&&&\\[-12pt]
$e^{2\pi i \gamma/g^2}$  & 1& 1 & 1  \\
&&&\\[-12pt]
$\phi_{ij}\equiv(\sigma, \phi)$  & 3& 1 & 0  \\
&&&\\[-12pt]
$\lambda_{ia}\equiv(\lambda, \psi_\phi)$  & 2& 2 &  0 \\
&&&\\[-12pt]
$u_a\equiv(u_+, u_-^*)$  & 1& 2 & 0  \\
&&&\\[-12pt]
$\psi_i\equiv(\psi_+,  \psi_-^*)$  & 2& 1 & 0 
\end{tabular}
\ee
\end{center}

The charges of the fields under the global abelian $U(1)_R \times U(1)_A \times U(1)_J$ and gauge $U(1)_a$ symmetries are given in \eqref{eq:thBchargescartan}.
\be\label{eq:thBchargescartan}
\begin{tabular}{c|cccc}
& $U(1)_R$ & $U(1)_A$ & $U(1)_J$ & $U(1)_a$ \\
\hline
&&&&\\[-12pt]
$u_+$ & 0 & 1 & 0 & 1 \\
&&&&\\[-12pt]
$u_-$ & 0 &1 & 0 & -1 \\
&&&&\\[-12pt]
$\psi_+$ & -1 & 1 & 0 & 1\\
&&&&\\[-12pt]
$\psi_-$ & -1 & 1 & 0 & -1\\
&&&& \\[-12pt]
\hline
&&&&\\[-12pt]
$e^{2\pi i \gamma/g^2}$  & 0& 0 & 1 & 0 \\
&&&&\\[-12pt]
$\sigma$  & 0& 0 & 0 & 0 \\
&&&&\\[-12pt]
$\phi$  & 2& -2 & 0 & 0 \\
&&&&\\[-12pt]
$\lambda$  & 1& 0 & 0 & 0 \\
&&&&\\[-12pt]
$\psi_\phi$  & 1& -2 & 0& 0  \\
\end{tabular}
\ee

The lagrangian of theory B is fixed by the symmetries and is nontrivial due to the interactions between the charged hypermultiplet and the emergent vector multiplet:
\be\label{eq:LB}
\mc L^{(B)}(\mc U, \mc V, \hat{\mc V}_J)= \mc L^{\mc V}(\mathcal V) + \mc L^{\mc H}(\mathcal U, \mathcal V)- \mc L_{BF}^{\mc N=4}(\mc V, \hat{\mc V}_J)\,.
\ee
The partition function of the theory is
\be
Z^{(B)}[\hat{\mc V}_J]= \int D \mc U\, D \mc V\,\exp\left(i \int d^3x\,\mc L^{(B)}(\mc U, \mc V,\hat{\mc V}_J)\right)\,.
\ee

Mirror symmetry states that the partition functions of theory A and B are the same:
\be\label{eq:N4}
Z^{(A)}[\hat{\mc V}_J]=Z^{(B)}[\hat{\mc V}_J]\,.
\ee
The global symmetries on both sides match\footnote{Note that we write global symmetries such that the matching is $SU(2)_{L, N} \leftrightarrow SU(2)_{L,N}$. This departs from the standard convention in mirror symmetry works where global symmetries are interchanged $SU(2)_L \leftrightarrow SU(2)_N$.}; the moduli space of theory A (the Higgs branch parametrized by $v_\pm$) maps to the moduli space of theory B -- the Coulomb branch parametrized by the scalars $\phi, \sigma, \gamma$.

\subsection{Deformations by $U(1)_A$ and $U(1)_R$ backgrounds}\label{subsec:defs}

We now consider an extension of mirror symmetry that includes $U(1)_A$ and $U(1)_R$ background deformations. This will be crucial for deriving the bosonization duality below. 

We first discuss the simpler case of global non-R-symmetries.
The basic observation is simple: both sides of the mirror pair have a conserved $U(1)_A$ current, so the partition function should agree also in the presence of a background gauge field that couples to the current. 
In fact, since $U(1)_A$ commutes with SUSY, we can introduce a $\mc N=2$ background vector superfield.

The background vector superfield $\hat V_A$ contains a scalar $\hat \sigma_A$, a gauge field $\hat A_{A}$, a gaugino $\hat \lambda_A$ and a D-term $\hat D_A$. Taking into account the charges of the elementary fields in \eqref{eq:thAchargescartan}, the $U(1)_A$ background deforms the lagrangian by
\be\label{eq:Abackground}
\mc L^{(A)}({\cal Q}, \hat{V}_A)= \int d^4 \theta \left( V_+^\dag e^{2 \hat  (V- \hat V_A)} V_++V_-^\dag e^{-2 (\hat  V+ \hat V_A)} V_- \right).
\ee

From \eqref{eq:thBchargescartan}, the background $U(1)_A$ couples to the chiral superfields $U_\pm$ and $\Phi$ in theory B as follows:
\begin{align}
\label{eq:Bbackground}
\mc L^{(B)}({\cal U}, {\cal V}, \hat{V}_A) = & \frac{1}{4g^2} \int d^2 \theta\,W_\alpha^2 +\text{h.c.} + {1 \over g^2} \int d^4 \theta \Phi^\dagger e^{- 4 \hat{V}_A} \Phi \cr
& + \int d^4 \theta \left( U_+^\dag e^{2   (V+ \hat V_A)} U_++U_-^\dag e^{-2  (V- \hat V_A)} U_-\right)-\frac{1}{2\pi} \int d^4 \theta\, V\, \hat \Sigma\,.
\end{align}
With the $U(1)_J$ and $U(1)_A$ backgrounds turned on, mirror symmetry implies
\be
Z^{(A)}[\hat V_J, \hat V_A]= Z^{(B)}[\hat V_J, \hat V_A]\,.
\ee

Finally, we consider a background superfield coupling to the $U(1)_R$ current $j_\mu^R$ and its SUSY completion. 
This is a bit more subtle than the previous case because $U(1)_R$ does not commute with SUSY. 
The superspace structure of the $U(1)_R$ symmetry multiplet and its linearized couplings have been recently worked out in~\cite{Komargodski:2010rb, Dumitrescu:2011iu} and their results may be used to map a SUSY background for $U(1)_R$ across the duality.

For this work, however, we only need the background gauge field $\hat A_{R}$ and its scalar partner $\hat \sigma_R$; these can be mapped across the duality without using the full superspace machinery. First, $ \hat A_{R}$ appears through the minimal coupling $ \hat A_{R, \mu} j^\mu_R$ (plus possible quadratic terms to ensure gauge invariance), and since $j^\mu_R$ is determined by the Noether procedure, it is straightforward to map $\hat A_{R}$ across the duality. On the other hand, the coupling to $\hat \sigma_R$ can be obtained by first working in a 3+1D theory with minimal coupling $\hat A_{R, m} j_R^m$ with $m=0, \ldots,3$, and then dimensionally reducing along $x^2$ and identifying $\sigma_R = i A_{R, m=2}$. Thus, $\hat \sigma_R$ couples to the extra-dimensional component of the current $j_R^{m=2}$. This agrees with the analysis in~\cite{Komargodski:2010rb, Dumitrescu:2011iu}.

\subsection{General mirror duality}\label{subsec:general}

In this way, we arrive at the general statement of mirror symmetry in the presence of backgrounds,
\be\label{eq:full-duality}
Z^{(A)}[\hat V_J, \hat V_A, \hat V_R]= Z^{(B)}[\hat V_J, \hat V_A, \hat V_R]\,.
\ee
%Formulas (\ref{eq:LA-backgrounds}) and (\ref{eq:LB-backgrounds}) in Appendix \ref{app:lagr} present in full detail the lagrangians in the presence of backgrounds.  
It is important to stress that \eqref{eq:full-duality} holds as long as the mass scales associated to the backgrounds are much smaller than the scale $g^2 \rightarrow \infty$ below which theory B flows to its interacting fixed point description.

Let us note one immediate consequence of (\ref{eq:full-duality}) that will be important below. Consider a point in the phase diagram of background couplings where some of the fermionic fields on both sides are massive. Integrating them out produces Chern-Simons terms for the background gauge fields as dictated by (\ref{eq:loopCS}). Some of the scalars can also condense, inducing Higgs masses for certain combinations of the background gauge fields -- these combinations disappear from the low energy theory. Then (\ref{eq:full-duality}) implies that the corresponding matrices of Chern-Simons levels $k_{MN}$, projected onto the subspace of massless fields, have to match between theory A and theory B. This is a direct consequence of the SUSY duality, but it also holds if SUSY is broken by some of the background D-terms, because even in this case, the partition functions must still be equal.
Our derivation of bosonization will make crucial use of this fact.

\section{Chiral mirror symmetry}\label{sec:chiralmirror}

As a step towards the bosonization relation \eqref{firstduality}, we first derive a chiral SUSY duality equating the theory of a free ${\cal N}=2$ chiral superfield to ${\cal N}=2$ SUSY QED3 with a single chiral superfield.
This is a particular case of a family of dualities dervied in \cite{Tong:2000ky}.
This is accomplished by turning on backgrounds $\hat \sigma_J$ and $\hat \sigma_A$. 
The effects of these perturbations are clear in theory A since it is free.
While the theory B description is strongly coupled, SUSY ensures that our analysis is reliable due to the absence of phase transitions as a function of the gauge coupling. 
%A similar approach was used in~\cite{DoreyTong2000, Tong:2000ky} to derive a duality with two flavors of chiral superfields. 
It will become clear that the chiral mirror duality provides a SUSY completion for \eqref{firstduality}.

\subsection{Chiral theory A}

In theory A,
let us turn on backgrounds 
\be
|\hat \sigma_A - \hat \sigma_J| \ll \hat \sigma_A \sim \hat \sigma_J.
\ee
More precisely, we write
\be
\hat \sigma_A= \hat \sigma_A^0+ \delta \hat \sigma_A\;,\;\hat \sigma_J = \hat \sigma_A^0+ \delta \hat{\sigma}_J\,,
\ee
with $|\delta \hat \sigma_{A,J}| \ll \hat \sigma^0_A$. 
Our goal is to derive an effective theory valid at energy scales $E \ll \hat \sigma_A^0$.
Within the effective theory, we will denote $\delta \hat{\sigma}_{A,J} = \hat{\sigma}_{A,J}$ for notational simplicity.

$V_-$ receives a large SUSY-preserving mass, while $V_+$ is light. 
Therefore, theory A reduces to the model of a free superfield with symmetries given in \eqref{eq:chiraltab1A}.
\be\label{eq:chiraltab1A}
\begin{tabular}{c|ccc}
& $U(1)_R$ & $U(1)_A$ & $U(1)_J$ \\
\hline
&&&\\[-12pt]
$V_+$ & 1 & -1 & 1  \\
&&&\\[-12pt]
\hline
&&&\\[-12pt]
$v_+$ & 1 & -1 & 1 \\
&&&\\[-12pt]
$\Psi_+$ & 0 & -1 & 1 \\
\end{tabular}
\ee
Note that in this theory the two global symmetries $U(1)_A$ and $U(1)_J$ act the same way on the dynamical fields.

At scales $E \ll \hat \sigma_A^0$, the effective description is
\begin{align}
\label{eq:LA-backgrounds}
\mc L^{(A)}_{\rm chiral} & = |D_{\hat{A}_J - \hat{A}_A + \hat{A}_R} v_+|^2 - \left((\hat \sigma_J - \hat \sigma_A + \hat \sigma_R)^2+\hat D_J - \hat D_A\right) |v_+|^2 \cr
& + i \bar \Psi_+ \slashed{D}_{\hat{A}_J - \hat{A}_A} \Psi_+ - (\hat \sigma_J - \hat \sigma_A) \bar \Psi_+ \Psi_+ + {1 \over 8 \pi} k^{(A)}_{MN} \hat{A}_M d \hat{A}_N,
\end{align}
%Finally, in the presence of backgrounds that are much smaller than the UV cutoff $\hat \sigma_A^0$, the Lagrangian deforms to
%\bea\label{eq:LA-backgrounds}
%\mc L_A &=& |\hat D_\mu v_+|^2+ i \bar \Psi_+ \not \! \!\hat D \Psi_+-\left((\hat \sigma_J - \hat \sigma_A + \hat \sigma_R)^2+\hat D_J - \hat D_A\right) |v_+|^2  - (\hat \sigma_J - \hat \sigma_A) \bar \Psi_+ \Psi_+ \nonumber\\
%&+&{1 \over 8 \pi} k^{(A)}_{MN} \hat{A}_M d \hat{A}_N
%\eea
where $\hat A_M = (\hat A_J, \hat A_A, \hat A_R)$ and the Chern-Simons ``K-matrix,"
\be\label{eq:kA1}
k^{(A)}_{MN}=\sgn(\hat \sigma_A^0)\, \left(\begin{matrix}-1 & -1 & 0 \\ -1 & -1 & 0 \\ 0 & 0 &0 \end{matrix} \right),
\ee 
comes from integrating out the $\Psi_-$ component of the superfield $V_-$.\footnote{The SUSY completion of the CS term will not play a role in what follows so it is not written.} 
Non-zero $\hat{D}_J$ or $\hat{D}_A$ break SUSY. 
The chiral theory in \eqref{eq:LA-backgrounds} is stable as long as scalar $v_+$ mass-squared is non-negative:
\begin{align}
\label{stabilitycondition}
m^2_{v_+} = (\hat \sigma_J - \hat \sigma_A + \hat \sigma_R)^2+\hat D_J - \hat D_A \geq 0.
\end{align}
Interactions must be included in order to study the regime of parameter space where $v_+$ is unstable.

%To end the discussion, let us note that starting from the $\mc N=4$ version and turning on $\hat \sigma_A \sim \hat \sigma_J$, a similar flow occurs and we are also left with the theory of a free superfield (\ref{eq:chiralLA}). However, the CS response matrix is different from (\ref{eq:kA1}) since in this case there is no integration over $\psi_M$. Furthermore, in the presence of SUSY breaking the theory remains exactly free, unlike (\ref{eq:WFint}).\footnote{More precisely, we expect corrections suppressed by $g^2$ instead of $\hat \sigma_A$, so the SUSY breaking relevant interactions are parametrically suppressed in the limit $g^2 \to \infty$ considered in this work.}

%%%%%%%%%%%%%%%%%%%%%%%%%%%%%%%%%%%%%%%%%%%%%%%
%%%%%%%%%%%%%%%%%%%%%%%%%%%%%%%%%%%%%%%%%%%%%%%
\subsection{Chiral theory B}

Consider next the effect of 
\be
\hat{\sigma}_A - \hat{\sigma}_J \ll \hat \sigma_A \sim \hat \sigma_J\ll g^2 \rightarrow \infty
\ee
in theory B. 
As before, we write the backgrounds as a large $\hat \sigma_A^0$ plus fluctuations that we denote by $\hat{\sigma}_{A,J}$ within the effective theory. 
Since the background axial mass $\hat{\sigma}_A$ appears in combination with the Coulomb branch scalars as $\sigma \pm \hat \sigma_A$,
% -- see Eq. \eqref{eq:LB-backgrounds} -- 
only one chiral multiplet can be light at a time, i.e., for a given value of $\sigma$. 
We will now show, in fact, that a SUSY-preserving vacuum for $\pm \hat{\sigma}_A > 0$ requires that the $U_\pm$ multiplet is massive.

To see this, let us integrate out both charged scalars $u_\pm$ under the assumption that both scalars are massive, $\sigma \pm \hat{\sigma}_A \neq 0$. This produces a new contribution to the potential that mixes the auxiliary D-field with $\sigma \pm \hat \sigma_A$. The terms that contribute to the effective potential are
\be\label{eq:VeffB1}
V_{\rm eff} = -\frac{1}{2g_{\rm eff}^2} D^2+ \frac{1}{2\pi} D \hat \sigma_J -\frac{1}{4\pi} D\,\left(| \hat \sigma_A + \sigma|-| \hat \sigma_A - \sigma| \right)\,.
\ee
The first term encodes the one-loop renormalization of the gauge coupling, which we discuss shortly. The second term is the FI term\footnote{A ``FI term" is one that is linear in $D$.} sourced by the background $\hat \sigma_J$, and the last term is produced by integrating out the massive $u_\pm$ scalars. This last effect may also be understood using SUSY: integrating out the $\psi_\pm$ fermion partners produces a mixed CS term between $a$ and $\hat A_{A}$, and its SUSY completion in (\ref{eq:LBF-general}) includes a term of the form $\pm D |\hat \sigma_A \pm \sigma|$.

A SUSY-preserving vacuum requires that the total FI term vanishes,
\be
\frac{1}{2} \left(| \hat \sigma_A + \sigma|-| \hat \sigma_A - \sigma| \right) - \hat \sigma_J=0\,.
\ee
For $0 < \hat \sigma_A = \hat \sigma_J$, this is accomplished by $\langle \sigma \rangle \geq |\sigma_A|$; or for $\langle \sigma \rangle < - |\hat{\sigma}_A|$ when $\hat{\sigma}_A = \hat{\sigma}_J < 0$. 
Now, since  $m_{u_\pm}^2 = (\sigma \pm \hat \sigma_A)^2$, we find that the massless superfield corresponds to $U_{-\sgn(\hat \sigma_A)}$ at $\sigma = \hat{\sigma}_A$, while $U_{{\rm sgn}(\hat{\sigma}_A)}$ is massive.

From \eqref{eq:thBchargescartan}, we see that $\Phi$ carries charge $-2$ under $U(1)_A$.
Thus, a non-zero $\hat{\sigma}_A$ results in a SUSY-preserving mass equal to $- 2 \hat{\sigma}_A$ for this multiplet.

To construct an effective theory for the remaining light modes, we specialize to the case $\sgn(\hat \sigma_A)>0$, and redefine the origin of the $\sigma$ field, 
\be
\tilde \sigma \equiv \sigma- \hat \sigma_A^0\,.
\ee
Since $U_+$ and $\Phi$ have tree-level masses, they may be integrated out.  
The resulting low-energy theory is a $\mc N=2$ chiral gauge theory with the matter content and charges assignments in \eqref{eq:thBchargescartanchiral}.
\be\label{eq:thBchargescartanchiral}
\begin{tabular}{c|cccc}
& $U(1)_R$ & $U(1)_A$ & $U(1)_J$ & $U(1)_a$ \\
\hline
&&&&\\[-12pt]
$U_-$ & 0 & 1 & 0 & 1\\
&&&&\\[-12pt]
$W_\alpha$ & 1 & 0 & 0 & 0\\
\hline
&&&&\\[-12pt]
$u_-$ & 0 &1 & 0 & -1 \\
&&&&\\[-12pt]
$\psi_-$ & -1 & 1 & 0 & -1\\
&&&& \\[-12pt]
$e^{2\pi i \gamma/g^2}$  & 0& 0 & 1 & 0 \\
&&&&\\[-12pt]
$\tilde{\sigma}$  & 0& 0 & 0 & 0 \\
&&&&\\[-12pt]
$\lambda$  & 1& 0 & 0 & 0 \\
&&&&\\[-12pt] \\
\end{tabular}
\ee
Note that in this effective theory, the $U(1)_A$ and $U(1)_a$ charges of the light fields are proportional. 
This is related to the redundancy between $U(1)_A$ and $U(1)_J$ in theory A.

Integrating out the massive $U_+$ has two additional effects. 
First, it produces a one-loop renormalization of the gauge coupling,
\be
\frac{1}{g_{\rm eff}^2}=\frac{1}{g^2}+ \frac{1}{8\pi |2 \hat \sigma_A^0+ \delta \hat \sigma_A + \t \sigma|}\,.
\ee
Thus, we may take the limit $g^2/\hat \sigma_A^0 \to \infty$ to obtain the effective gauge coupling $g_{\rm eff}^2 \approx 16 \pi \hat \sigma_A^0$. 
The scale $\hat \sigma_A^0$ works as the UV cutoff of the effective theory.

The second effect is the generation of a level-1/2 ${\cal N}=2$ Chern-Simons term of the form in Eq. \eqref{N2CS}.
This ${\cal N}=2$ Chern-Simons term may be decomposed into a Chern-Simons term at level $k=1/2$ for the dynamical gauge field, a shift of the FI term as discussed above, and a mass proportional to $-g_{\rm eff}^2$ for the gaugino $\lambda$.

Integrating out the D-term and the gaugino at the classical level, we conclude that below the cutoff $\hat \sigma_A^0$, the effective description of theory B is captured by
\bea\label{eq:LBchiral}
\mc L^{(B)}_{\rm eff}&=&\frac{1}{8\pi} a d a +\frac{1}{2g_{\rm eff}^2}(\partial \t\sigma)^2+ | D_{-a} u_-|^2+ i \bar \psi_- \slashed{D}_{-a} \psi_- - \t \sigma^2 |u_-|^2 \nonumber\\
&-& \frac{g_{eff}^2}{2}\left(|u_-|^2+\frac{\t \sigma}{4\pi}\right)^2-8 \pi \bar \psi_- \psi_- |u_-|^2\,.
\eea
Thus, we have shown that by deforming $\mc N=4$ mirror symmetry, we arrive at the following duality:
\be\label{eq:chiral-duality}
\text{free chiral superfield}\;V_+ \;\leftrightarrow\; U(1)_{{1 \over 2}}\;\text{with charged superfield}\;U_-\;\text{and neutral scalar}\;\t \sigma.
\ee
This has the following interesting consequence. At the origin $\tilde{\sigma} = 0$,
%, the field $\t \sigma$ is massive and can be integrated out. For energies $E \ll \hat \sigma_A^0$, 
the effective lagrangian becomes
\be
\mc L^{(B)}_{\rm eff}= \frac{1}{8\pi} a d a + | D_{-a} u_-|^2+ i \bar \psi_- \slashed{D}_{-a} \psi_- -16 \pi^2 |u_-|^6 -8 \pi \bar \psi_- \psi_- |u_-|^2\,.
\ee
The order-one interactions are classically marginal with vanishing one-loop beta functions \cite{Avdeev:1991za}. 
Furthermore, it was argued in \cite{Gaiotto:2007qi} that this fixed point survives to all orders in perturbation theory.  Therefore, the theory flows to a superconformal field theory (SCFT).   
The duality implies that this interacting SCFT admits a description in terms of a free chiral superfield (which nevertheless is highly nonlocal in terms of the original variables).

We now include background fields within the effective theory with magnitude much smaller than the (effective) UV cutoff $\hat \sigma_A^0$. 
Keeping the auxiliary $D$-field explicit, we find
\begin{align}
\label{eq:LBback}
\mc L^{(B)}_{\rm chiral}&=\frac{1}{2g_{\rm eff}^2}\left((\partial \t\sigma)^2+D^2\right)+ | D_{-a + \hat{A}_A} u_-|^2+ \bar \psi_- i \slashed{D}_{- a + \hat{A}_A - \hat{A}_R} \psi_- -( (\t \sigma- \hat \sigma_A)^2-D+\hat D_A) |u_-|^2 \nonumber\\
&-8 \pi \bar \psi_- \psi_- |u_-|^2-(-\t \sigma+ \hat \sigma_A- \hat \sigma_R) \bar \psi_- \psi_-+\frac{1}{8\pi} (a + \hat A_{A}- \hat A_{R}) d (a+ \hat A_{A}- \hat A_{R})\nonumber\\
&+ \frac{1}{4\pi} ( \t\sigma+ \hat \sigma_A ) (D+ \hat D_A)-\frac{1}{8\pi} \hat A_{R} d \hat A_{R} - {1 \over 8 \pi} (2 \hat{A}_A - \hat{A}_R) d (2\hat{A}_A - \hat{A}_R) \cr
& -\frac{1}{2\pi}\left(\hat A_{J} d a+ \hat D_J \t\sigma+ D \hat \sigma_J\right)\,.
\end{align}
As before, there is a slight abuse of notation here: the background values $\hat \sigma_A$ and $\hat \sigma_J$ are small deviations from $\hat \sigma_A^0$ that were turned on in the UV. 
The last terms of the second line and first terms of the third line contain the Chern-Simons terms generated by integrating out $\psi_+$, the gaugino, and $\psi_\phi$. The remaining terms in the fourth line are the BF couplings to the background $U(1)_J$ fields. 
It is now straightforward to integrate out $D$, yielding the effective potential for the scalar fields,
\be
V_{\rm eff}^{\rm chiral}= (\t \sigma^2+ \hat D_A) |u_-|^2+\frac{g_{\rm eff}^2}{2} \left(|u_-|^2+\frac{1}{4\pi}(\t \sigma+ \hat \sigma_A - 2 \hat \sigma_J) \right)^2\,.
\ee

%%%%%%%%%%%%%%%%%%%%%%%%%%%%%%%%%%%%%%%
%%%%%%%%%%%%%%%%%%%%%%%%%%%%%%%%%%%%%%%
\subsection{Moduli space and ``charge attachment''}

By deforming mirror symmetry, we have obtained the new SUSY duality (\ref{eq:chiral-duality}). We will now perform various checks on this, beginning with a matching of the moduli space of both theories.

When $\hat \sigma_A = \hat \sigma_J$ and $\hat \sigma_R= \hat D_A = \hat D_J=0$,
theory A has a massless field $v_+$, that is charged under the three $U(1)$ global symmetries. 
In the absence of SUSY breaking deformations, the vacuum expectation value (VEV) of $v_+$ parameterizes an exact modulus. An expectation value $\langle v_+ \rangle$ breaks one linear combination of the global symmetries and manifests itself as a Higgs mass,
\be\label{eq:HiggsA}
\mc L^{(A)} \supset -|\langle v_+ \rangle|^2 (\hat A_{J}- \hat A_{A}+ \hat A_{R})^2\,.
\ee

To see the corresponding effect in theory B, let us focus on the dynamics of $\tilde \sigma$ for $\hat \sigma_R= \hat D_A = \hat D_J=0$. 
When $\tilde{\sigma}$ has a nonzero VEV, $u_-$ and $\psi_-$ are massive, and integrating them out produces a one-loop correction similar to (\ref{eq:VeffB1}):
\be
\tilde{V}_{\rm eff}=- \frac{1}{2 \tilde{g}_{\rm eff}^2} D^2- \frac{1}{4\pi} D \left(\hat \sigma_A+ \t \sigma-|\hat \sigma_A- \t \sigma|-2 \hat \sigma_J \right)
\ee
and
\be
\frac{1}{\tilde{g}_{eff}^2}= \frac{1}{8\pi} \left(\frac{1}{2 \hat \sigma_A^0+ \hat \sigma_A + \t \sigma}+ \frac{1}{|\hat \sigma_A - \t \sigma|} \right)\,.
\ee
In this last expression, we have distinguished explicitly the large UV value $\hat \sigma_A^0$ from the small fluctuation $\hat \sigma_A$ in order to avoid confusion. 
Note that in the IR limit $\hat \sigma_A^0 \to \infty$, the new renormalized gauge coupling becomes $\tilde{g}_{\rm eff}^2= 8 \pi |\hat \sigma_A - \t \sigma|$. The condition to have a SUSY vacuum is the vanishing of the D-term,
\be
\hat \sigma_A+ \t \sigma-|\hat \sigma_A- \t \sigma|-2 \hat \sigma_J=0\,.
\ee
For $0 < \hat \sigma_A = \hat \sigma_J$, we then find an exactly flat direction $\tilde \sigma> \hat \sigma_A$, while $\t \sigma< \hat \sigma_A$ is lifted. Furthermore, when $\t \sigma> \hat \sigma_A$, integrating out $\psi_-$ generates a Chern-Simons contribution for $a$ that cancels the corresponding term in (\ref{eq:LBback}). As a result, we find an additional massless real scalar from the dual photon, and hence the moduli space has complex dimension one. This is in agreement with the moduli space of theory A.

We should also understand how global charges match along the moduli space. For this, consider a nonzero VEV $\langle \t \sigma \rangle>\hat \sigma_A >0$. Then $\psi_-$ is massive (with a sign opposite to that of $\psi_+$), and the Chern-Simons terms produced upon integrating out $\psi_+, \psi_-, \lambda,$ and $\psi_\phi$ combine to give
\be
\mc L_{CS}^{(B)}=- \frac{1}{2\pi} a d (\hat A_{J}- \hat A_{A}+ \hat A_{R})- \frac{1}{8\pi} \hat A_{R}d \hat A_{R} - {1 \over 8 \pi} (2 \hat{A}_A - \hat{A}_R) d (2 \hat{A}_A - \hat{A}_R)\,.
\ee
Recalling that $a$ is dynamical, its equation of motion sets
\be\label{eq:HiggsB}
\hat A_{J}- \hat A_{A}+ \hat A_{R}=0\,.
\ee
In other words, this combination of fields is set to zero in the low energy theory. 
But this is precisely the same effect as the Higgs mechanism (\ref{eq:HiggsA}) in theory A. 
The Chern-Simons and BF couplings ``attach'' global charges to $\tilde \sigma$ in a way that matches the charges of $v_+$ and result in the Higgsing of the same linear combination of fields given on the left-hand side of (\ref{eq:HiggsB}). 
This ``charge attachment mechanism'' is essentially the dual of ``flux attachment" \cite{Fradkinbook} and is likewise implemented by Chern-Simons couplings; it was found in a string theory context in \cite{MaldacenaMooreSeiberg2001}.

%%%%%%%%%%%%%%%%%%%%%%%%%%%%%%%%%%%%%%%
%%%%%%%%%%%%%%%%%%%%%%%%%%%%%%%%%%%%%%%
\subsection{Massive SUSY-preserving deformations}\label{subsec:susy-massive}

The chiral duality can be further tested by turning on the background $\hat \sigma_A$ and $\hat \sigma_J$, which produce SUSY-preserving masses.

From (\ref{eq:LA-backgrounds}), theory A becomes gapped, with both $v_+$ and $\Psi_+$ acquiring mass $\hat \sigma_J-\hat \sigma_A$. 
The resulting gapped theory is characterized by the Chern-Simons response:
\be
\mc L^{(A)}_{CS}={1 \over 8 \pi} k^{(A)}_{MN} \hat{A}_M d \hat{A}_N
\ee
with
\be\label{eq:kA}
k^{(A)}_{MN}=\left(\begin{matrix}-1 & -1 & 0 \\ -1 & -1 & 0 \\ 0 & 0 &0 \end{matrix} \right)+\sgn(\hat \sigma_J-\hat\sigma_A)\left(\begin{matrix}1 & -1 & 0 \\ -1 & 1 & 0 \\ 0 & 0 &0 \end{matrix} \right)\,.
\ee
The first term comes from (\ref{eq:kA1}), while the second term is produced by integrating out $\Psi_+$.

The dynamics in theory B is somewhat more complicated, since the stabilization involves also quantum effects. 
%There are four possibilities for $(\hat \sigma_A, \hat \sigma_J)$ depending on whether they are positive or negative and and on which has larger absolute value. For the purpose of matching (\ref{eq:kA}), it will be sufficient to consider positive backgrounds and $\hat \sigma_A>\hat \sigma_J$ or $\hat \sigma_A<\hat \sigma_J$.
Let us consider the case $\hat \sigma_A > \hat \sigma_J>0$ first. Anticipating that we will find a minimum for $\t \sigma$ away from $\hat \sigma_A$, we integrate out the massive $u_-$ field to find the D-term:
\be
D=-\frac{g_{\rm eff}^2}{4\pi} (\hat \sigma_A+\t \sigma-|\hat \sigma_A-\t \sigma|-2 \hat \sigma_J)\,;
\ee
see the discussion around (\ref{eq:VeffB1}). This has a SUSY-preserving vacuum  at $\langle\t \sigma\rangle = \hat \sigma_J$. This is the unique global vacuum, i.e., there is no SUSY vacuum with $\langle u_-\rangle \neq 0$.
The fermion $\psi_-$ acquires then a mass $\hat \sigma_A - \hat{\sigma}_J$, and integrating it out we 
obtain the contribution 
\begin{align}
\delta {\cal L}^{(B)}_{\rm CS} = {1 \over 8 \pi} \Big(- a + A_{A} - A_{R}\Big) d \Big(- a + A_{A} - A_{R}\Big)\,.
\end{align}
Adding this to the contribution of the topological term found previously in (\ref{eq:LBback}) and the BF term between $\hat{A}_J$ and $a$, we obtain:
\be
\mc L^{(B)}_{CS}=\frac{1}{4\pi} \left[(a d a - 2 a d \hat A_{J}) - 2\hat A_{A} d \hat A_{A}\right]\,.
\ee
Integrating out the dynamical gauge field $a$ reproduces the K-matrix in Eq. (\ref{eq:kA}) when $\hat \sigma_A>\hat \sigma_J>0$.

The stabilization mechanism is different if $\hat \sigma_J > \hat \sigma_A>0$.
We will self-consistently find that $\langle \t \sigma \rangle = \hat \sigma_A$. 
Since this corresponds to the point where the tree-level mass for $u_-$ vanishes, let us return to the value of the D-term before integrating out $u_-$:
\be
D=-g_{\rm eff}^2 \left(|u_-|^2+\frac{1}{4\pi}(\t \sigma+\hat \sigma_A-2 \hat \sigma_J) \right)\,.
\ee
The effective scalar potential is
\be
V_{\rm eff}=(\hat \sigma_A-\t \sigma)^2 |u_-|^2+\frac{g_{eff}^2}{2} \left(|u_-|^2+\frac{1}{4\pi}(\t \sigma+\hat \sigma_A-2 \hat \sigma_J) \right)^2\,.
\ee
The SUSY-preserving minimum lies at
\be
\langle \t \sigma \rangle = \hat \sigma_A\;,\;|\langle u_-\rangle|^2=\frac{\hat \sigma_J-\hat \sigma_A}{2\pi}\,.
\ee
The expectation value for $u_-$ has two effects. First, it produces a Higgs mass for the combination $a - \hat A_{A}$, and so at low energies we should set
\be
a - \hat A_{A}=0\,.
\ee 
Furthermore, from the quartic coupling $|u_-|^2 \bar \psi_- \psi_-$ (obtained by integrating out the gaugino), $\psi_-$ becomes massive and integrating it out produces a level-1/2 Chern-Simons term for the combination $-a+ \hat A_{A}- \hat A_{R}$. 
Adding these to the topological terms in (\ref{eq:LBback}), we obtain
\be
\mc L_{CS}^{(B)}=-\frac{1}{2\pi} \hat A_{J} d \hat A_{A}\,,
\ee
thus matching the K-matrix in Eq. (\ref{eq:kA}) for $\hat \sigma_J > \hat \sigma_A>0$.

This concludes the analysis of the nearby massive phases obtained by SUSY-preserving deformation in the chiral duality. To end, let us write the general Chern-Simons responses that must match as a result of the duality.\footnote{This statement is slightly imprecise: for a given theory, only the fractional part of the level of the Chern-Simons response is well defined \cite{ClossetDumitrescuFestucciaKomargodskiSeiberg}; however, the difference in this response across a phase transition is physical.} 
In theory A with arbitrary $\Psi_+$ and $v_+$ masses, we have the response,
\begin{align}
\label{eq:LAresponse}
\mc L^{(A)}_{\rm CS}&=\frac{1}{8\pi}\sgn(m_{\Psi_+}) (\hat A_{J}- \hat A_{A}) d (\hat A_{J}- \hat A_{A})- \Theta(-m_{v_+}^2) (\hat A_{J}- \hat A_{A}+ \hat A_{R})^2 \cr
& +  \frac{1}{8\pi} \left[ -(\hat A_{J}+ \hat A_{A}) d (\hat A_{J} + \hat A_{A}) \right],
\end{align}
where the term proportional to the step function with $\Theta(x>0) = 1$ and $\Theta(x<0) = 0$ is short-hand for the effect from Higgsing. 
In theory B,
\begin{align}
\label{eq:LBresponse}
\mc L^{(B)}_{\rm CS}&=\frac{1}{8\pi}\sgn(m_{\psi_-}) (-a+\hat A_{A}- \hat A_{R}) d (-a+\hat A_{A}- \hat A_{R})-\Theta(- m_{u_-}^2) (-a+ \hat A_{A})^2 \cr
&+\frac{1}{8\pi} \left[ (a+\hat A_{A}- \hat A_{R}) d (a+\hat A_{A}- \hat A_{R}) - 4 \hat{A}_A d \hat{A}_A + 4 \hat{A}_A d \hat{A}_R - 2\hat A_{R} d \hat A_{R}-4 \hat{A}_{J} d a \right]\,.\cr
\end{align}
This response will be crucial for our study of bosonization in the next section.

%%%%%%%%%%%%%%%%%%%%%%%%%%%%%%%%%%%%%%%%%%%%%%%%%%%%%
%%%%%%%%%%%%%%%%%%%%%%%%%%%%%%%%%%%%%%%%%%%%%%%%%%%%%
%%%%%%%%%%%%%%%%%%%%%%%%%%%%%%%%%%%%%%%%%%%%%%%%%%%%%
%%%%%%%%%%%%%%%%%%%%%%%%%%%%%%%%%%%%%%%%%%%%%%%%%%%%%
\section{Free Dirac fermion $\leftrightarrow$ scalar QED3}
\label{bosonizationarguments}

In this section, we use the SUSY duality of \S \ref{sec:chiralmirror} as a starting point to obtain the duality \eqref{firstduality} between a Dirac fermion and scalar QED3. 
The basic strategy is to break SUSY in a controlled way using a background $\hat D_J$ perturbation; we will then argue that for set $\hat D_J$ and varying $\hat \sigma_A$ and $\hat \sigma_J$, the SUSY duality deforms to \eqref{firstduality}.

%\subsection{Free Dirac fermion $\leftrightarrow$ scalar QED3}\label{subsec:duality2}

\subsection{Theory A: free Dirac fermion}

The demonstration of \eqref{firstduality} proceeds by assuming the hierarchy
\be\label{hierarchy}
(\hat \sigma_A-\hat \sigma_J)^2\ll \hat D_J\ll (\hat \sigma_A^0)^2\,.
\ee

From the quadratic lagrangian in (\ref{eq:LA-backgrounds}), $v_+$ is heavy and may be integrated out, but $\Psi_+$ remains as a light field. $\Psi_+$ is massless at the critical point and obtains a mass $m_{\Psi_+} = \hat{\sigma}_J - \hat{\sigma}_A$ away from the critical point.
We refer to these two massive phases as the $\hat{\sigma}_J - \hat{\sigma}_A > 0$ and $\hat{\sigma}_J - \hat{\sigma}_A < 0$ phases.
Neither of these two phases break the $U(1)_R \times U(1)_A \times U(1)_J$ global symmetry.

The critical theory has the effective description,
\begin{align}
\label{freedirac}
\hat{{\cal L}}^{(A)}_{\rm Dirac} = \bar{\Psi}_+ i \slashed{D}_{\hat{A}_J - \hat{A}_A} \Psi_+ - m_{\Psi_+} \bar{\Psi}_+ \Psi_+ + {k^{\rm crit}_{MN} \over 8 \pi} \hat{A}_M d \hat{A}_N
\end{align}
with 
\begin{align}
k^{\rm crit}_{MN} = \begin{pmatrix} -1 & -1 & 0 \cr -1 & -1 & 0 \cr 0 & 0 & 0 \end{pmatrix}.
\end{align}
Setting $\hat{A}_A = \hat{A}_R = 0$ and renaming $\Psi_+ = \Psi$ and $\hat{A}_J = \hat{A}$, we find the left-hand side of \eqref{firstduality} at the critical point $m_{\Psi_+} = 0$.

The topological response away from the critical point is given by (\ref{eq:kA}),
\be
\mc L^{(A)}_{CS}={1 \over 8 \pi} k^{(A)}_{MN} \hat{A}_M \,d\hat{A}_N
\ee
with
\be\label{eq:kA2}
k^{(A)}_{MN}=\left(\begin{matrix}-1 & -1 & 0 \\ -1 & -1 & 0 \\ 0 & 0 &0 \end{matrix} \right)+\sgn(\hat \sigma_J-\hat\sigma_A)\left(\begin{matrix}1 & -1 & 0 \\ -1 & 1 & 0 \\ 0 & 0 &0 \end{matrix} \right)\,.
\ee
\begin{figure}[h!]
  \centering
\includegraphics[width=.8\linewidth]{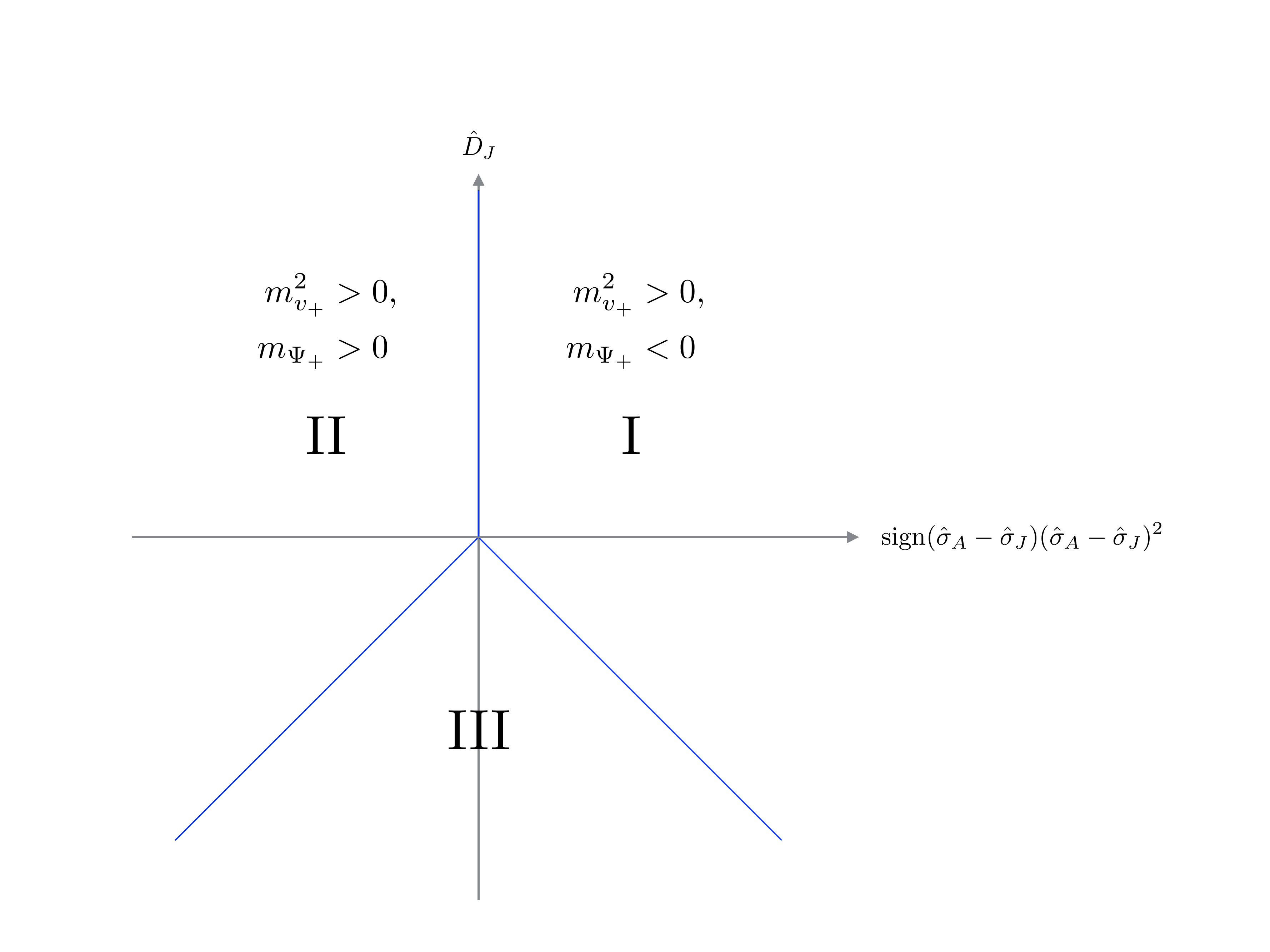}\\
\caption{{\bf Phase diagram of theory A.} Phases I-III are separated by second order critical points (indicated by the solid blue line). Setting $\hat{A}_A = 0$, the transition at $\hat{\sigma}_A = \hat{\sigma}_J$ represents the point across which the Chern-Simons level for $\hat{A}_J$ changes by unity. The horizontal axis at $\hat{D}_J = 0$ is described by the SUSY chiral theory A, while the $\hat{D}_J>0$ line is controlled by the free fermion lagrangian in Eq. \eqref{freedirac}. Phase III is unstable because $m^2_{v_+} < 0$ -- see Eq. \eqref{stabilitycondition} -- and there are no interactions to stabilize the broken-symmetry vacuum.}
\label{pd1}
\end{figure}
We thus arrive at the phase diagram in Fig. \ref{pd1}.

\subsection{Theory B: scalar QED3}

Let us now consider the effects of the background deformations in Eq. \eqref{hierarchy} on the theory B side of the dual chiral pair of \S \ref{sec:chiralmirror}. 
Duality implies that there is a {\it single} critical point as $\hat{\sigma}_A - \hat{\sigma}_J$ is varied about zero (within the regime of parameter variations we consider) in theory B.
We now show how to uniquely constrain what field must become light at the critical point by using the topological response (\ref{eq:LBresponse}) to the background gauge fields in the nearby massive phases. 
It is important to stress that the matching of topological responses is a consequence of the SUSY duality, and remains valid as long as the SUSY breaking scale is below the UV cutoff $\hat \sigma_A^0$ of the chiral mirrors.

%As in \S \ref{fqed3section}, we use duality and the existence of a single critical point (within the regime of parameter variations we consider) to argue that the free Dirac theory has a dual description in terms of scalar QED3.
Away from the critical point at $\hat{\sigma}_A - \hat{\sigma}_J = 0$, the theory is massive and we may parameterize via effective masses the topological response lagrangian of theory B as in (\ref{eq:LBresponse}). 
Matching with (\ref{eq:kA2}) uniquely determines
\bea
m_{\psi_-}(\hat \sigma_J < \hat \sigma_A)&>&0\;,\; m_{u_-}^2(\hat \sigma_J < \hat \sigma_A)>0 \nonumber\\
m_{\psi_-}(\hat \sigma_J > \hat \sigma_A)&>&0\;,\; m_{u_-}^2(\hat \sigma_J > \hat \sigma_A)<0\,.
\eea
In particular, the sign of the fermion mass is fixed by requiring that there be no pure BF coupling between $a$ and background gauge fields, and hence no global symmetry breaking. We conclude that $u_-$ is massless at the critical point, while $\psi_-$ is gapped in this region of the phase diagram and can be integrated out.

The last remaining field to consider is $\tilde{\sigma}$.
%When studying fermionic QED3 in \S \ref{fqed3section}, we argued that the sign of the effective mass-squared of $\tilde{\sigma}$ changes sign across the critical point from the relation between the $\langle |v_+|^2 \rangle$ and $\langle \tilde{\sigma} \rangle$.
Recall that in the SUSY theory discussed in the previous section, there was an identification of the moduli spaces of theory A and theory B which are (partially) parameterized by $\langle |v_+|^2 \rangle$ and $\langle \tilde{\sigma} \rangle$.
Because there is no breaking of the $U(1)_R \times U(1)_A \times U(1)_J$ global symmetry in either phase -- $\langle |v_+|^2 \rangle$ vanishes -- we do not expect $\langle \tilde{\sigma} \rangle$ to be non-zero.
The simplest scenario, consistent with broken SUSY, is for $\tilde{\sigma}$ to have a positive mass-squared across the transition.
Consequently, we have the effective description near the critical point,
\begin{align}
\label{sQED3}
{\cal L}_{\rm sQED3}^{(B)} = & |D_{-a + A_A} u_-|^2 - m^2_{u_-} |u_-|^2 - \lambda_{u_-} |u_-|^4 + {1 \over 4 \pi} a d a - {1 \over 2 \pi} \hat{A}_J d a - {1 \over 4 \pi} \hat{A}_A d \hat{A}_A.
\end{align}
The effective mass-squared $m^2_{u_-} = 0$ at the critical point and the quartic $|u_-|^4$ interaction obtains from integrating out massive fields.
Setting $\hat{A}_A = \hat{A}_R = 0$ and renaming $u_- = \varphi$ and $\hat{A}_J = \hat{A}$, we recover the right-hand side of \eqref{firstduality}.

\begin{figure}[h!]
  \centering
\includegraphics[width=.8\linewidth]{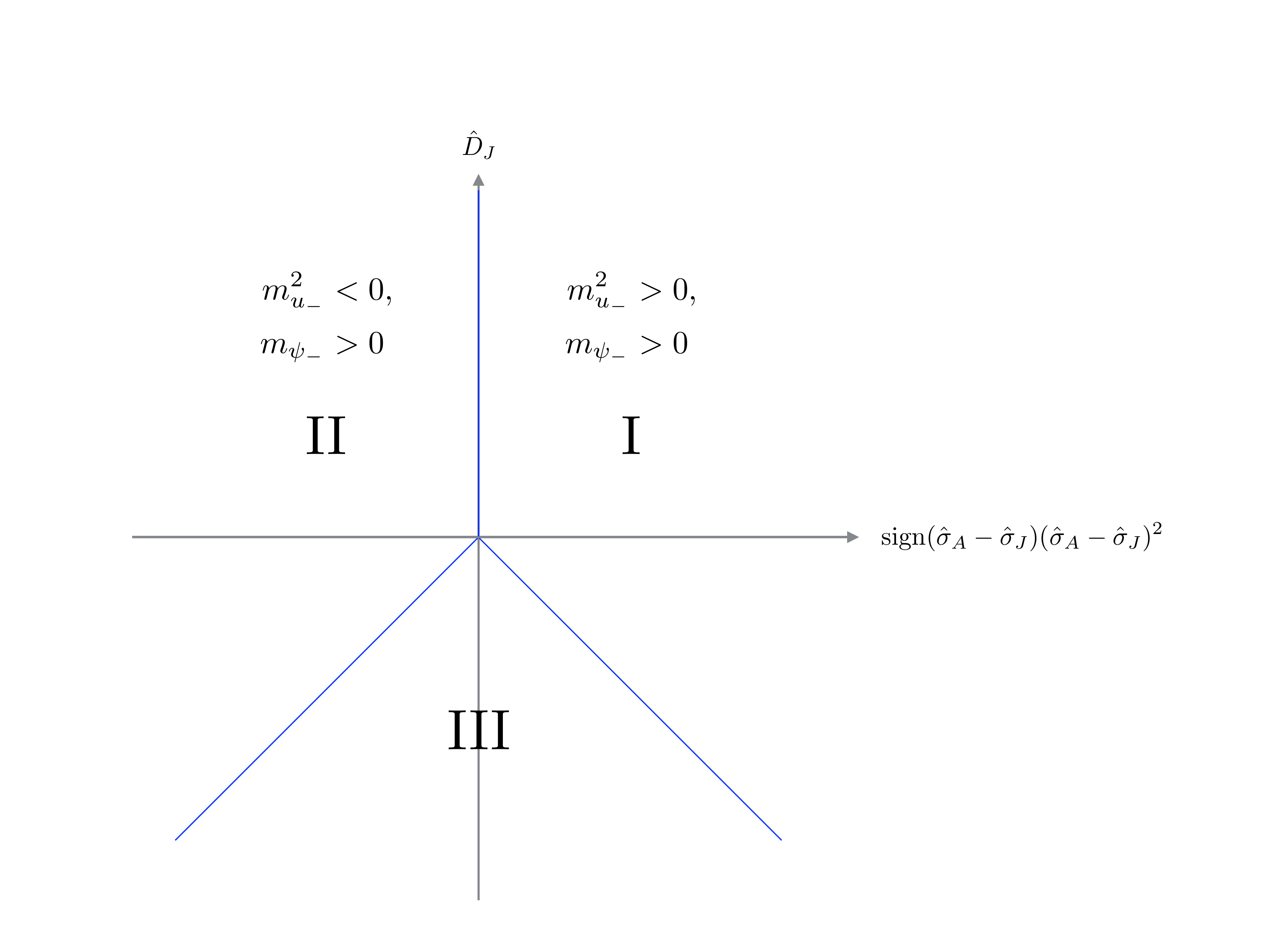}\\
\caption{{\bf Phase diagram of theory B.} Phases I-III are separated by second order critical points (indicated by the solid blue line). Setting $\hat{A}_A = 0$, the transition at $\hat{\sigma}_A = \hat{\sigma}_J$ represents the point across which the Chern-Simons level for $\hat{A}_J$ changes by unity. The horizontal axis at $\hat{D}_J = 0$ is described by the SUSY chiral theory B, while the $\hat{D}_J>0$ line is controlled by the lagrangian in Eq. \eqref{sQED3}. Phase III cannot be accessed within our framework.}
\label{pd2}
\end{figure}

As required by duality, the phase diagram in Fig. \ref{pd2} for theory B matches that of theory A in Fig. \ref{pd1}.
Setting $\hat{A}_A = 0$ and identifying $\hat{A}_J$ with electromagnetism, we have an effective description for an integer quantum Hall plateau transition: the point across which the Chern-Simons level for $\hat{A}_J$ changes by unity.
The two massive phases are determined by the sign of the fermion mass in theory A, while they are realized via an order-disorder transition of the scalar in the QED3 theory B.

As recent work \cite{Karch:2016sxi, Seiberg:2016gmd, 2016arXiv160601912M, HsinSeiberg2016} has shown, if \eqref{firstduality} is assumed, various additional dualities can be found upon the application of a modular transformation \cite{WittenSL2Z2003, LeighPetkouSL2Z2003}.
For instance, \eqref{secondduality} is the ${\cal S}$ transform of \eqref{firstduality}.

%%%%%%%%%%%%%%%%%%%%%%%%%%%%%%%%%%%%%%%%%%%%%%%%%%%%%
%%%%%%%%%%%%%%%%%%%%%%%%%%%%%%%%%%%%%%%%%%%%%%%%%%%%%
%%%%%%%%%%%%%%%%%%%%%%%%%%%%%%%%%%%%%%%%%%%%%%%%%%%%%
\section{Conclusions and future directions}\label{sec:concl}

In this work, we have shown how the 2+1-dimensional bosonization duality in \eqref{firstduality} -- relating the theory of a free Dirac fermion to scalar QED3 -- may be obtained by deforming supersymmetric ${\cal N}=4$ mirror symmetry. 
We first derived a ``chiral" supersymmetric version of the duality in which the theory of a free superfield is dual to supersymmetric QED3 with a single charged superfield. We then broke supersymmetry using a background D-term and showed that the chiral duality flows to the bosonization duality. 
As mentioned in the introduction,
modular transformations relate \eqref{firstduality} to a second bosonization duality between the Wilson-Fisher fixed point and fermionic QED3 \eqref{secondduality}, as well Peskin-Dasgupta-Halperin duality (\ref{eq:PDH}) and the topological completion (\ref{eq:son}) of the fermion/fermion conjecture in \cite{Son2015, WangSenthilfirst2015, MetlitskiVishwanath2016}.

We end by listing future directions that would be interesting to pursue. First, our methods may be applied to mirror symmetry when the number of flavors of chiral superfields $N_f>1$ and to certain quiver gauge theories. This would lead to a rich structure of bosonization dualities, which we hope to analyze in the future.
%It would be quite interesting to know if our response arguments are as uniquely constraining in such higher-$N_f$ examples.

The duality in \eqref{firstduality} represents the transition point between two massive phases -- see the dual phase diagrams in Figs. \ref{pd1} and \ref{pd2} -- where the level of the Chern-Simons term for a background $U(1)$ gauge field changes by unity across the transition. 
Thus, the critical point describes an integer quantum Hall plateau transition. 
At $\hat{D}_J = 0$, this critical point enjoys ${\cal N}=2$ supersymmetry, while supersymmetry is broken when $\hat{D}_J > 0$.
It would be interesting to include disorder in this system: How do the critical properties depend on whether or not supersymmetry is preserved?
Does unbroken supersymmetry provide an advantage to calculating disorder-averaged quantities, similar the technique introduced in \cite{ParisiSourlas1979} for non-supersymmetric systems?

The phase diagram Fig. \ref{pd1} includes a regime in which there is no stable vacuum: the mass-squared of the boson $v_+$ becomes negative.
Higher-order, e.g., quartic, interactions are necessary to stabilize the vacuum. 
Upon their successful inclusion, it would be interesting to explore the negative mass-squared regime where mirror symmetry suggests additional dualities may be lurking. 
%mention O(4) WF <-> two flavor fermionic QED3

It would be quite interesting to take the non-relativistic limit of \eqref{firstduality} and \eqref{secondduality}. 
Such a limit would presumably make contact with the ``flux attachment" procedures that have been used to study various strongly correlated systems \cite{Fradkinbook}. This has been addressed very recently for nonabelian dualities in \cite{Radicevic:2016wqn}.

A beautiful derivation of the fermion/fermion conjecture \cite{Son2015, WangSenthilfirst2015, MetlitskiVishwanath2016}, as stated by (\ref{eq:son}) when $b$ is integrated out, was provided in \cite{MrossAliceaMotrunichexplicitderivation2016}.
This derivation makes use of an anisotropic limit of the left-hand side of (\ref{eq:son}) in which one spatial direction is discretized while the second remains continuous. 
A clever change of variables enables a rewriting of this coupled-wire Dirac system in terms of fermionic QED3 with a single flavor.
Is there a refinement of this derivation that makes the $\nu=1/2$ bosonic Laughlin sector of the topological completion of the fermion/fermion conjecture in (\ref{eq:son}) manifest?

Duality implies matching 3-sphere free energy \cite{KapustinWillettYaakov, Jafferis2012, JafferisKlebanovPufuSafdi2011} or disk entanglement \cite{MyersSinha2011, CasiniHuertaMyers2011, CasiniHuerta2012} for certain dual pairs.
Localization can be applied to mirror pairs deformed by non-zero FI D-terms and mass parameters that preserve ${\cal N}=2$ supersymmetry \cite{KapustinWillettYaakov}.
It would be interesting to perform this test on the chiral mirrors of \S \ref{sec:chiralmirror}.
A related question concerns the matching of the first-quantized wave functions for dual pairs. 
%mishmash paper?

%global phase diagram; other fixed points

\section*{Acknowledgments}

We thank E. Dyer, E. Fradkin, S. Raghu, and D. Tong for helpful discussions.
We are grateful to A. Karch, J. Murugan, N. Seiberg, T. Senthil, D. Tong, and C. Wang for very helpful comments and correspondence on an early draft of this manuscript.
This research was supported in part by the National Science Foundation under grants NSF PHY-1316699 (S.K.) and NSF PHY-11-25915 (M.M.), the John Templeton Foundation (S.K. and M.M.), and Conicet PIP-11220110100752 (G.T.). 
M.M. is grateful for the generous hospitality of the Aspen Center for Physics, which is supported by National Science Foundation grant NSF PHY-1066293, where part of this work was performed and the Kavli Institute for Theoretical Physics in Santa Barbara where this work was completed.

%%%%%%%%%%%%%%%%%%%%%%%%%%%%%%%%%%%%%%%%%%%%%%%%%%%%%
%%%%%%%%%%%%%%%%%%%%%%%%%%%%%%%%%%%%%%%%%%%%%%%%%%%%%
%%%%%%%%%%%%%%%%%%%%%%%%%%%%%%%%%%%%%%%%%%%%%%%%%%%%%
\appendix

\section{Spinors and superspace in 2+1 dimensions}\label{app:superspace}

Let us review how $\mc N=2$ superspace in 2+1 dimensions follows from $\mc N=1$ in 3+1 dimensions. For this, it is convenient to adopt the conventions of Wess and Bagger~\cite{Wess:1992cp}, though we do not use them in the main part of the text. A Weyl fermion in 3+1 dimensions becomes a Dirac fermion in 2+1 dimensions, and indices are raised and lowered using the antisymmetric tensor $\epsilon^{\alpha \beta}= i \sigma_2$, $\epsilon^{12}=1$.\footnote{For conventions similar to the ones we use in this appendix, see \cite{Ivanov:1991fn, Benna:2008zy,Dumitrescu:2011iu}.}

If $\psi_\alpha$ is a Weyl fermion, indices are raised and lowered in terms of the antisymmetric tensor $\epsilon^{\alpha \beta}=i \sigma^2$, $\epsilon^{12}=1$. In particular, the following conventions for contracting indices are very helpful,
\bea\label{eq:mult}
\psi \chi &=& \psi^\alpha \chi_\alpha = (\epsilon^{\alpha \beta} \psi_\beta) \chi_\alpha\nonumber\\
\psi^\dag \chi^\dag &=& \psi^\dag_{\dot \alpha} \chi^{\dag \dot \alpha}=\psi^\dag_{\dot \alpha}(\epsilon^{\dot \beta \dot \alpha} \chi^\dag_{\dot \beta})\,.
\eea
With these conventions, typical lagrangian terms for a Weyl fermion include
\be
L = i \psi^\dag \bar \sigma^m D_m \psi - \frac{M}{2}(\psi \psi+ \psi^\dag \psi^\dag)\;,\;D_m = \partial_m - i g A_m\,.
\ee

We perform the dimensional reduction along $x^2$, and would like to identify $A_2$ as giving rise to a mass term in the 2+1D theory:
\be\label{eq:32}
\psi^\dag \bar \sigma^m A_m \psi \Big|_{3+1D}= \bar \psi \gamma^\mu A_\mu + M \bar \psi \psi\Big|_{2+1D}\,.
\ee
In 2+1D we also use the convention (\ref{eq:mult}) for products of spinors, though this time there are no dotted indices $\theta^\alpha= \epsilon^{\alpha \beta}\theta_\beta\;,\;(\theta^\dag)^\alpha= \theta_\beta^\dag \epsilon^{\beta \alpha}$ and
\bea\label{eq:spinor-metric}
\psi^\dag \psi&\equiv & \psi^\dag_\alpha \psi^\alpha= \psi^\dag_\alpha \epsilon^{\alpha \beta } \psi_\beta\\
\psi \psi^\dag & \equiv & \psi^\alpha \psi^\dag_\alpha=\epsilon^{\alpha \beta}\psi_\beta \psi^\dag_\alpha =- \psi^\dag \psi\nonumber\,.
\eea
From this and (\ref{eq:32}), we deduce
$(\gamma^\mu)^{\alpha \beta}= (\mathbf 1, -\sigma^1, -\sigma^3)$
and $A_2=i M$. The more standard representation where products are taken with the Kronecker delta is $(\gamma^\mu)^\alpha_{\;\beta}=(\gamma^\mu)^{\alpha \sigma}\epsilon_{\sigma \beta}=(- i \sigma^2, \sigma^3, \sigma^1)$.

It is now straightforward to obtain the $\mc N=2$ superspace in 2+1D by starting from the $\mc N=1$ superspace in 3+1D, dropping the $x^2$ dependence, and using the above conventions for spinors and gamma matrices. We find, from the formulas in Wess-Bagger, that the superspace derivatives, chiral and vector superfields are given by
\bea\label{eq:superspace}
D_\alpha &=& \frac{\partial}{\partial \theta_\alpha}+ i (\gamma^\mu \bar \theta)_\alpha \partial_\mu \nonumber\\
\Phi&=& \phi(y) + \sqrt{2} \theta \psi(y) + \theta^2 F(y)\;,\;y^\mu= x^\mu+ i \theta \gamma^\mu \bar \theta\nonumber\\
V&=& -i \theta \bar \theta\,\sigma+ \theta \gamma^\mu \bar \theta A_\mu + i \theta^2 \bar \theta \bar \lambda- i \bar \theta^2 \theta \lambda+ \frac{1}{2}\theta^2 \bar \theta^2 D\,.
\eea
$\sigma$, the scalar component of a 2+1D vector multiplet, now appears simply from the dimensional reduction of the 3+1D vector multiplet, $A_2= -i \sigma$. 

We may similarly translate the lagrangians in superspace:
\bea
\int d^4 \theta\,\Phi^\dag e^{2q V} \Phi&=&-|D_\mu \phi|^2+ i \bar \psi \not \! \! D \psi-(q \sigma)^2 |\phi|^2-q \sigma \bar \psi \psi-i q (\phi^* \lambda \psi - \phi \bar \lambda \bar \psi)+ q D |\phi|^2;\nonumber\\
\frac{1}{4} \int d^2 \theta\,W_\alpha^2 +\text{h.c.}&=&-\frac{1}{4} F_{\mu\nu}^2-\frac{1}{2}(\partial \sigma)^2+i \bar \lambda \not \!\partial \lambda+\frac{1}{2}D^2\,,
\eea
where $D_\mu = \partial_\mu -i q A_\mu$. 
The equations in \S \ref{sec:overview} follow by changing the spacetime signature and the representation for the gamma matrices.

\bibliography{QHE.bib}{}
\bibliographystyle{utphys}
\end{document}